\documentstyle[11pt,newpasp,twoside,psfig]{article}
\def\fe{\langle Fe \rangle} 

\def\edcomment#1{\iffalse\marginpar{\raggedright\sl#1\/}\else\relax\fi}
\reversemarginpar

\begin{document}
\title{Mg$_2$-$\sigma$ in Early-Type Galaxies and 
Spiral Bulges}
\author{Cristina Chiappini}
\affil{Osservatorio di Trieste, Via G. B. Tiepolo 11, Trieste 34131, Italy}
\author{Paulo Pellegrini, Charles Rit\'e, Marcio Maia, Ricardo Ogando, B. Ramos}
\affil{$^*$Observat\'orio Nacional, Rio de Janeiro, Brazil}
\author{Ricardo P. Schiavon, Christopher N.A. Willmer$^*$, USA}
\affil{UCO/Lick Observatory, Univ. of California, Santa Cruz, USA}
\author{Luiz da Costa$^*$}
\affil{European Southern Observatory, Garching, Germany}
\author{Mariangela Bernardi}
\affil{Dept. of Astronomy and Astrophysics, Univ. of Chicago, USA}
\author{Maria V. Alonso}
\affil{Observatorio Astr\'onomico de Cordoba, Cordoba, Argentina}
\author{Gary Wegner}
\affil{Dept. of Physics and Astronomy, Dartmouth College, Hanover, USA}

\begin{abstract}
We analyze new measurements of the Mg$_2$ central line strength index
and velocity dispersion ($\sigma$) for the galaxies of the ENEAR survey. 
The observations are now complete (da Costa et al. 2000) and the sample 
contains 1223 early-type galaxies. We also analyze the line strength indices for a 
sample of 95 spiral bulges (from Sa to Sbc). 
For the early-type galaxies we find: {\it i)}
that the Mg$_2$-$\sigma$ relation for Es and
S0s are nearly the same, with both populations showing comparable
scatter, and {\it ii)} a marginal difference in the slope of the
Mg$_2$ and $\sigma$ relation for cluster and field early-type galaxies.
However, we suggest that before interpreting such a difference in the 
framework of a mass-metallicity relation, it is important to take into
account the effects of rotation in the Mg$_2$-$\sigma$ relation. 
Our preliminary results indicate that once the rotation effects
are minimized by choosing a sample containing only slow rotators,
the Mg$2$-$\sigma$ relation is similar
both for isolated and clustered galaxies. More data on rotational
velocities of early-type galaxies are needed to confirm this result.
For spiral bulges, we find that their locus in the 
Mg$_2$-$\sigma$ plane lies always below the one 
occupied by early-type galaxies.
\end{abstract}

\section{Introduction}

In this work we focus our attention on the Mg$_2$-$\sigma$ relation presented
by early-type galaxies (e.g. Terlevich et al. 1981) and its possible dependence
on the environment as suggested by some authors.
The goal is to investigate the existence of this difference and, if real, its
possible cause. We also show a comparison of early-type 
galaxies and spiral bulges in the Mg$_2$-$\sigma$ diagram. While there is growing
evidence that at least the well studied bulges (Milky Way and M31) are old (Renzini
1999 and references therein), it is extremely important to investigate if bulges
share the same chemical properties of E+S0s. For instance, Jablonka et al. (1996)
and Idiart et al. (1996) found that bulges (from Sas to Scs) follow the same 
Mg$_2$-$\sigma$ relation displayed by E+S0s.
In this work we address the above questions by means of an 
analysis of Lick index measurements
of integrated spectra of E+S0s and bulges which has proven to be important
for studies of the formation and evolution of galaxies. 
The data will be presented elsewhere (Rit\'e et al. 2001,
Wegner et al. 2001).  

\section{The data}

The samples used in this analysis come from two surveys conducted
by our group at Observat\'orio Nacional (Rio de Janeiro) and collaborators. 
Early-type galaxies are primarily from the recently completed ENEAR redshift
distance survey of da Costa et al. (2000), while the spirals are taken from 
SSRS2 sample (da Costa et al. 1998). 

The ENEAR project 
assembled spectroscopic and photometric data for early-type galaxies
brighter than m$_B$=14.5, covering the whole sky. 
From these, 1223 galaxies have both spectroscopic and photometric
data and were used in this analysis.
Spectroscopic data for these galaxies 
yielded radial velocities, central velocity dispersions 
and several indices in the Lick system. 
Photometric data in the R band yielded total magnitudes, intensity 
profiles, bulge and disk characteristic scales, disc-to-bulge ratios, isophotal 
shape parameters and a characteristic diameter D$_n$ (Dressler et al. 1987).
Moreover, by establishing an appropriate Dn-$\sigma$ relation in the R band
true distances were obtained for most of the objects
in this sample.
The ENEAR sample was split between field and group/cluster galaxies by using the 
membership assignment as described in
Bernardi et al. (1998) (hereafter B98). 
In order to 
simplify the analysis we split the sample in two categories: isolated galaxies
and clustered galaxies. The latter class combines groups and clusters and 
the two-fold separation basically represents low and high density
environments. We note that as compared with the analysis of B98
there is an improvement in the group catalog used resulting in a more 
realistic distribution with environment namely, a larger number of clustered
early-type galaxies.

The sample of spirals comes from the Southern Sky Redshift Survey
(SSR2 da  Costa et al. 1998) down to a limiting magnitude $m_B$ = 15.5.
Here we only considered spirals earlier than Sbc and galaxies with
central intensity ratios bulge/disk larger than 2.
In the case of spirals, the integrated light at large radii is dominated
by the disk, making an aperture correction procedure very uncertain. Therefore,
spectral extraction of spiral bulges was done by varying the projected
aperture so as to preserve a fixed metric diameter of 1.19h$^{-1}$kpc. 
A major concern in measuring the spectral properties of bulges of spiral
galaxies is the effect of contamination of the integrated spectra by disk-light.
We carried out many tests to estimate the level of this contamination (see
Pellegrini et al. 2001b) and we concluded that the disk contribution
in our integrated spectra is negligible.

\section{Early-Type Galaxies and the Mg$_2$-$\sigma$ relation}

\subsection{Mg$_2$-$\sigma$ for field and cluster galaxies: slope difference}

The ENEAR database yields the Mg$_2$-$\sigma$ relation displayed in Figure 1a,
where the data are split between isolated and clustered galaxies using the membership
assignment mentioned before. 
In contrast with B98 (see below)
we find a statistically significant (even though marginal) difference in the Mg$_2$-$\sigma$
fit between the cluster and field samples (Figure 1a).
As the sample includes both ellipticals and S0s, we also examine the possibility 
that morphology could be the cause of this difference. Splitting the total sample into 
ellipticals and S0s and performing a least square fit to the Mg$_2$-$\sigma$
relation (see Figure 1b), we find that the
resulting relations are essentially the same.
Before interpreting the origin of this result and its and implications 
for the scenario of galaxy formation we need to consider further parameters
that can affect the Mg$_2$-$\sigma$ relation and as a consequence its interpretation
as a mass-metallicity relation (section 3.3).

\subsection{Comparison with Bernardi et al. 1998}

Instead of looking for a difference in the slope of the Mg$_2$-$\sigma$
relation for clustered and field galaxies (as we did in section 3.1), 
B98 assumed a unique slope for early-type galaxies (obtained from the whole
sample) and looked for a zero-point difference among galaxies
belonging to different environments. 
These authors find a zero-point difference in the Mg$_2$-$\sigma$ relation between
cluster and field galaxies of $\Delta$Mg$_2$ = 0.007 $\pm$ 0.002.
Under the assumption of a fix slope, we also find a zero-point
difference of 0.010 $\pm$ 0.002, in agreement with B98.
A similar result was reported by Jorgensen (1997) who
found $\Delta$Mg$_2$ = 0.009 $\pm$ 0.002.

B98 concluded that the zero-point
difference of 0.007 $\pm$ 0.002 found in the Mg$_2$ index would imply a 
difference in age of less than 1Gyr, field galaxies being younger. 
The important implication of this result is that 
if early-type galaxies were assembled through mergers, this process
would have had to be fast, taking place at high redshifts.

\subsection{Mg$_2$-$\sigma$ and the effects of rotation}

Before discussing the origin of the result reported in section 3.1, 
lets consider the following 
arguments. It is usually assumed that the correlation of Mg$_2$ with velocity 
dispersion is equivalent to the more fundamental one: the mass-metallicity relation (Larson 1974).
However, restricting ourselves to the velocity
dispersion as a representation of galaxy masses, two relevant parameters
are not being taken into account namely the characteristic size
and the rotational velocity of the galaxies. Indeed, if a rotating galaxy is in equilibrium, 
the Virial Theorem enables
us to derive its total mass, by measuring some characteristic parameters 
related to internal dynamics and size as shown bellow:

\begin{equation}
M_{vir} = k (3 \langle \sigma^2 \rangle +  \langle V_{rot}^2 \rangle ) R_e = 
k' \langle \sigma^2 \rangle R_e (1 + (1/3)(\langle V_{rot}^2 \rangle/\sigma^2))
\end{equation}

\noindent
where the quantities $\langle \sigma^2 \rangle$ and $\langle V_{rot}^2 \rangle$
are the effective values of the mean square line-of-sight velocity dispersion and
the mean square rotation velocity, respectively, and $R_e$ is the effective radius.
After some simplifying assumptions, equation (1) can be written as (see Prugniel
and Simien 1994):

\begin{equation}
M_{vir} = k' \langle \sigma^2 \rangle R_e (1 + 0.81(V_{rot}/\sigma)^2)
\end{equation}

\noindent
where k' $\simeq$ 2$\times$10$^6$ if M$_{vir}$ is in units of solar masses, $\sigma$ and 
V$_{rot}$ are in km/s and R$_e$ in kpc. From the above expression it is clear that the Mg$_2$-$\sigma$
relation
represents a mass-metallicity relation only for slow rotating objects for which the approximation
$M = k' \sigma^2 R_e$ is valid. Figure 1c shows this effect. In this figure, we plot all early-type
galaxies from the ENEAR database for which rotational velocity measurements are available 
in the literature
(HYPERCAT database by Prugniel and collaborators and Rix et al. 1999). 
The figure also shows the different fits obtained
for objects with $V_{rot} < $50km/s (solid line) and $V_{rot} > $50km/s (dashed line).
The slope difference found among fast and slow rotators is larger than the difference
found when comparing different environments.
In this figure (dotted lines) an estimate of the 
locus occupied by galaxies with $V_{rot}$ = 50, 100 and 200km/s 
is also shown (see Pellegrini et al. 2001a). 
These curves represent an estimate of the effects of rotation
in the Mg$_2$-$\sigma$ relation and the main consequence is to produce a larger scatter
at lower $\sigma$ value (depending on details of the normalization adopted those dotted curves
can be shifted up and down, so what is important here is their overall effect).
As it is clearly seen from this figure, rotation moves data points to the right
in the Mg$_2$-$\sigma$ diagram. When splitting the objects with $V_{rot} <$ 50 km/s into
clustered and field galaxies, the Mg$_2$-$\sigma$ relation does not seem to depend 
on environment, although this result should be confirmed using a larger sample.

In summary, although we find a difference in the slope of the Mg$_2$-$\sigma$ relation
among clustered and field galaxies, when considering the completed ENEAR sample, 
this result cannot be interpreted in the framework of a mass-metallicity relation unless the 
effects due to rotation are taken into account.

\begin{figure}
\centerline{\psfig{figure=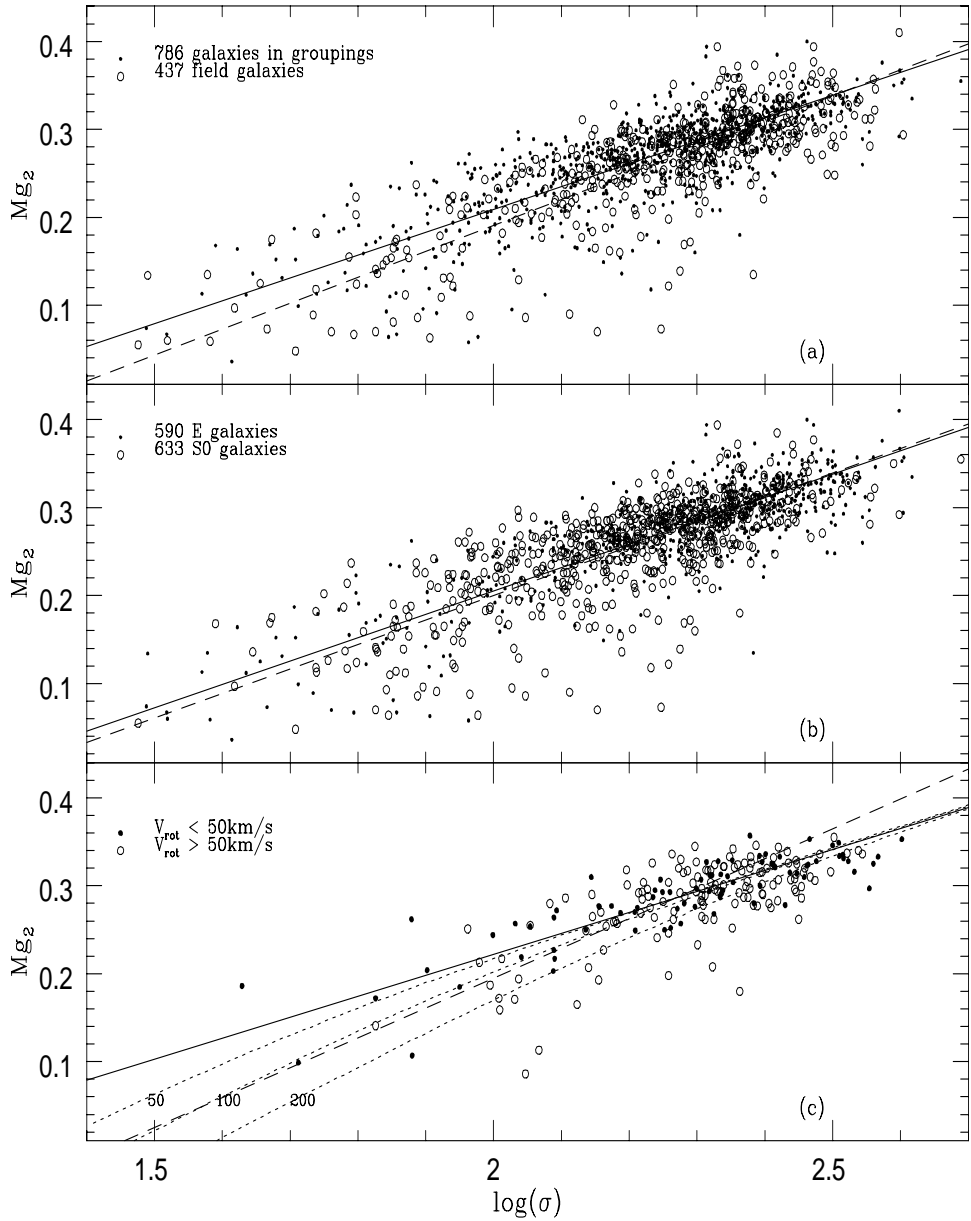,width=14cm,height=18cm}}
\vspace{-1.0truecm}
\caption{Mg$_2$-$\sigma$ relation obtained from ENEAR database: a) for cluster E+S0s (dots) and
field E+S0s (open circles).
resampling). Orthogonal least-square mean fits with bootstrap
resampling are shown for galaxies in groups/clusters (solid line) 
and in the field (dashed line); b) the same as a) but now instead of dividing the objects according
to the environment, we slit the sample into ellipticals 
(fit shown by a solid line) and S0s (fit shown by a 
dashed line); c) same as in a) but only for galaxies that have measured rotational velocities. The 
sample was divided into slow (solid line fit) and fast rotators (dashed line fit). The dotted lines
show the locus of the solid line in case a rotational velocity
of 50, 100 and 200 km/s was considered.} 
\end{figure}

\section{Early-type galaxies vs Spiral Bulges and the $Mg_2$-$\sigma$ relation }

Figure 2 shows the fit for early-type galaxies with $V_{rot}<$ 50km/s as
well as the curves obtained when including rotation superposed on the distribution of
early-type galaxies (panel a) and spiral bulges (panel b).
In Figure 2b the open symbols
correspond to the bulges in our sample and the crosses are the data by
Prugniel et al. (2001). The agreement between both samples is remarkable.
This figure shows that although bulges follow the same trend of 
increasing Mg$_2$ for increasing $\sigma$ as E+S0s, a clear shift to the 
lower right region is seen. Moreover, the slope of the Mg$_2$-$\sigma$ relation
for bulges is steeper than the one found in early-type galaxies.
Only massive bulges seem to have Mg$_2$ line strenghts comparable with 
E+S0 galaxies.


\begin{figure}
\centerline{\psfig{figure=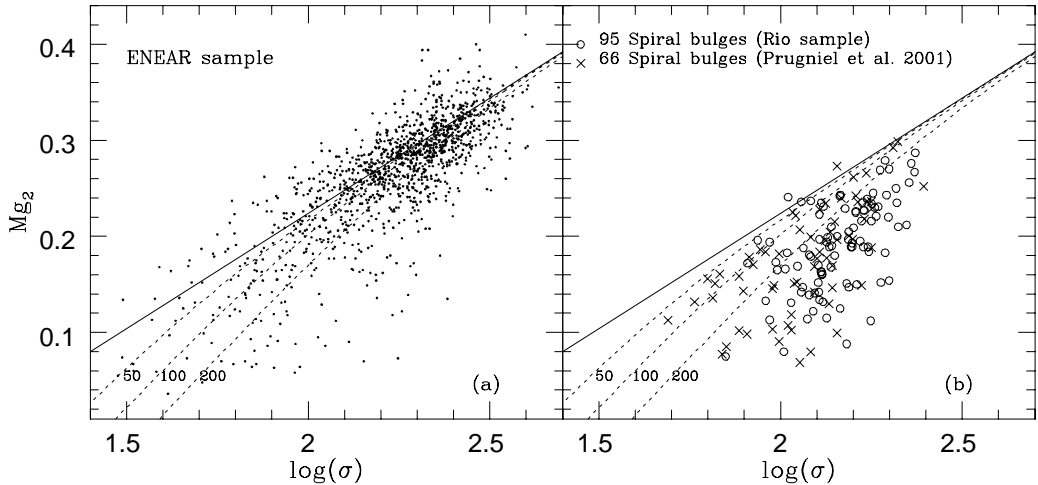,height=15cm}}
\vspace{-8.0truecm}
\caption{Mg$_2$-$\sigma$ relation traced by a) early-type
galaxies b) by the bulge data sample present here (open
circles) and Prugniel et al. (2001) sample (crosses). 
Lines are the same as shown in figure 1c.}
\end{figure}

Our findings are in close agreement with the recent results
reported by Prugniel et al. (2001). Both results are in
contrast with earlier conclusions of Jablonka
et al. (1996) based on a much smaller sample of spiral galaxies
than the one considered here. The amplitude of the effect is 
considerably larger than the small offset to lower
line strengths for the bulges noted by some authors (e.g.
Wyse et al. 1997 inspecting the same data) and qualitatively
similar to the results of Balcells and Peletier (1994) who find
that bulges and ellipticals of the same luminosity do not have
the same color, with bulges being bluer.

Part of this difference can be artificially caused in an
Mg$_2$-$\sigma$ plane since rotation is improperly
ignored (bulges are known to be fast rotators).
However another reason for this difference
can be related to differences in their chemical
evolution and by the contamination of the bulge
population by younger stars (secular evolution and 
later infall). By analyzing preliminary results based on the
iron index, $\fe$, we find that although early-type galaxies do not show
a significant relation between $\fe$ and $\log\sigma$,
bulges do show a clear correlation between those quantities
(exactly the same result found by Prugniel et al. 2001) showing
that bulges and early-type galaxies show different chemical properties.

\section{Discussion}

In view of the above results, we suggest a possible scenario for the 
formation of spheroids to be tested. We suggest that
the formation of early-type galaxies occurs 
from a gas cloud in which gravity is acting to produce its collapse and primordial angular 
momentum of the cloud tends to retain gas in the outer parts of the system,
which can later fall back into the system. 
A rapid initial collapse (involving all
protogalaxy mass or a large fraction of its inner portions) will produce a spheroid (older stars)
and later infall of gas may contribute to contaminate such an old population
by a younger one. If a galaxy resides in a low density environment, free from 
gas stripping, this contamination will be
more efficient and those galaxies would now show a low Mg$_2$
index (as a consequence of the addition of gas and the formation
of a non-dominant younger population). 
On the other hand, if the galaxy is in a high
density environment the gas in the outer parts of high rotation objects
would be removed by the frequent encounters, 
inhibiting the infall that could cause further star formation.
Moreover, galaxies with low rotational velocities would have had a more
efficient collapse and hence their evolution would be less dependent
on the environment.
To test the above scenario one would need to verify if: {\it i)} 
the Mg$_2$-$\sigma$ relation for slow rotators
is the same in field and clusters; {\it ii)} the 
Mg$_2$-$\sigma$ relation for fast rotators
is very different in field and clusters; {\it iii)}
fast rotators in the field show larger H$\beta$ index values
as compared with fast rotators in clusters.

The observational differences concerning line strength-velocity
dispersion relation of bulges and ellipticals can also be understood
in this same scenario, bulges being the 
extreme in the sequence of increasing rotation.
However
in this case the differences would also be a result
of different chemical evolution and later contamination by a younger population.
This is likely to happen as we expect
some degree of cross-talk between disks and bulges.

\end{document}